\documentclass[10pt,superscriptaddress,tightenlines,twocolumn,amsmath,amssymb,aps, pra]{revtex4-2}
\usepackage{amssymb}
\usepackage{amsmath}
\usepackage{amsfonts}
\usepackage{graphicx}
\usepackage{makecell}
\usepackage{hyperref}
\usepackage{makecell}
\usepackage{color}%
\usepackage{lmodern}
\usepackage{lineno}

\providecommand{\U}[1]{\protect\rule{.1in}{.1in}}
\hypersetup{colorlinks=true, linkcolor=blue, citecolor=red, urlcolor=blue}

\begin{document}

\title{Simultaneous All-versus-Nothing Refutation of Local Realism and Noncontextuality by a Single System}

\author{Min-Gang Zhou}
\affiliation{National Laboratory of Solid State Microstructures and School of Physics, Collaborative Innovation Center of Advanced Microstrucstures, Nanjing University, Nanjing 210093, China}
\affiliation{Department of Physics and Beijing Key Laboratory of Opto-electronic Functional Materials and Micro-nano Devices, Key Laboratory of Quantum State Construction and Manipulation (Ministry of Education), Renmin University of China, Beijing 100872, China}
\author{Hua-Lei Yin}\email{hlyin@ruc.edu.cn}
\affiliation{Department of Physics and Beijing Key Laboratory of Opto-electronic Functional Materials and Micro-nano Devices, Key Laboratory of Quantum State Construction and Manipulation (Ministry of Education), Renmin University of China, Beijing 100872, China}
\affiliation{National Laboratory of Solid State Microstructures and School of Physics, Collaborative Innovation Center of Advanced Microstrucstures, Nanjing University, Nanjing 210093, China}
\author{Zeng-Bing Chen}\email{zbchen@nju.edu.cn}
\affiliation{National Laboratory of Solid State Microstructures and School of Physics, Collaborative Innovation Center of Advanced Microstrucstures, Nanjing University, Nanjing 210093, China}

\begin{abstract}
The quantum realms of nonlocality and contextuality are delineated by Bell's theorem and the Kochen-Specker theorem, respectively, embodying phenomena that surpass the explanatory capacities of classical theories. These realms hold transformative potential for the fields of information and computing technology. In this study, we unveil a ``all-versus-nothing" proof that concurrently illustrates the veracity of these two seminal theorems, fostering a more nuanced comprehension of the intricate relationship intertwining quantum nonlocality and contextuality. Leveraging the capabilities of three singlet pairs and a Greenberger-Horne-Zeilinger state analyzer, our proof not only substantiates the conflict between quantum mechanics and hidden-variable theories from another perspective, but can also be readily verifiable utilizing the existing linear optics technology. 
\bigskip

\end{abstract}

\maketitle
\section{INTRODUCTION}

Despite the fact that quantum mechanics (QM) is by far the most successful physical theory,
its counterintuitive nature remains a conceptual barrier to understanding
its foundations. Einstein,
Podolsky and Rosen (EPR) discovered one of the physics' most famous paradoxes~\cite{einstein1935can} among those early efforts on quantum foundations: According to the assumption of local realism,
a maximally entangled two-party system appears to allow one to infer
definite values of complementary variables (e.g., position and momentum) by
using perfect correlations, which contradicts QM. Several hidden-variable theories were inspired by the EPR paradox. To deny hidden variables in QM, two major
theorems, Bell's theorem, \cite{bell1964physics} and the Kochen-Specker (KS)
theorem \cite{kochen1967,kupczynski2024quantum,simon2000,simon2001,cabello2008,yu,context-nature,chenjl}, have been developed. Based on EPR's notion of local
realism, the former denies local hidden variables (LHV), whereas the latter poses a more serious challenge to hidden variables.
The KS theorem states that it is impossible to assign definite values to each
of not all commutative observables in an individual system, a phenomenon known as quantum 
contextuality. This means that even in the absence of the locality condition, the value of an observable is dependent
on the experimental context, thus ruling out noncontextual hidden variables (NCHV).
In addition, Klyachko \textit{et al}. derived inequalities~\cite{Klyachko,Kunjwal} that are valid for NCHV, but
violated for individual quantum systems.

The original Bell's theorem argument is strongly statistical,
and perfect two-qubit correlations would not violate Bell's inequalities. To
highlight the contradiction between QM and LHV for definite predictions, the
``Bell's theorem without inequalities'', 
also known as the Greenberg-Horne-Zeilinger (GHZ) theorem, was developed, which has been demonstrated for the case of
three~\cite{greenberger1989bell,greenberger1990bell,mermin1990pt,pan2000experimental} or
even two~\cite{cabello2001all,PhysRevLett.95.210401,chen2003all,yang2005all} spacelike
separated observers. Because this proof without inequalities is based on perfect
correlations, it directly refutes EPR's notion of local realism, implying a stronger refutation of local realism.

Quantum nonlocality (i.e., violations of Bell's inequalities) and quantum contextuality both empower certain quantum information tasks~\cite{context-nature,Cabello2021}, according to quantum information science. The search for possible relationships between the two nonclassical phenomena follows naturally. Since the works of
Stairs~\cite{Stairs}, Heywood and Redhead~\cite{HeywoodandRedhead},
Mermin~\cite{mermin1990simple}, and R.D. Gill~\cite{van2005statistical,acin2005optimal}, recent years have seen significant advances
on the interiguing connections~\cite{Cabello2021} between quantum contextuality and 
nonlocality in terms of inequalities, such as converting local contextuality into 
nonlocality~\cite{Cabello2010,Cabello2021prl,guo2016}, simultaneously observing contextuality 
and nonlocality~\cite{guo2018,xue2022}, and a trade-off relation between nonlocality
and contextuality~\cite{Cabello2014,guo-pra,xue2016}.

In this study, we present an demonstration that both Bell's theorem and the KS theorem, inherently devoid of inequalities, can be illustrated harmoniously within a single system. Drawing inspiration from Mermin's proof~\cite{mermin1990simple} centered on three-qubit Pauli operators for the KS theorem and the ``all-versus-nothing" proof of Bell's theorem~\cite{chen2003all,chen2005avn,ghz2005}, our proposal can be viewed as a natural extension of these theorems. It simultaneously rejects both local and non-contextual hidden variables, adopting an all-versus-nothing approach. Within the contemporary landscape of rapid advancements, our proof furnishes compelling evidence supporting the correlation between quantum contextuality and nonlocality, offering a fresh lens to perceive their interrelationship.

\section{THEORY}

\begin{figure}[ptb]
\includegraphics[width=\columnwidth]{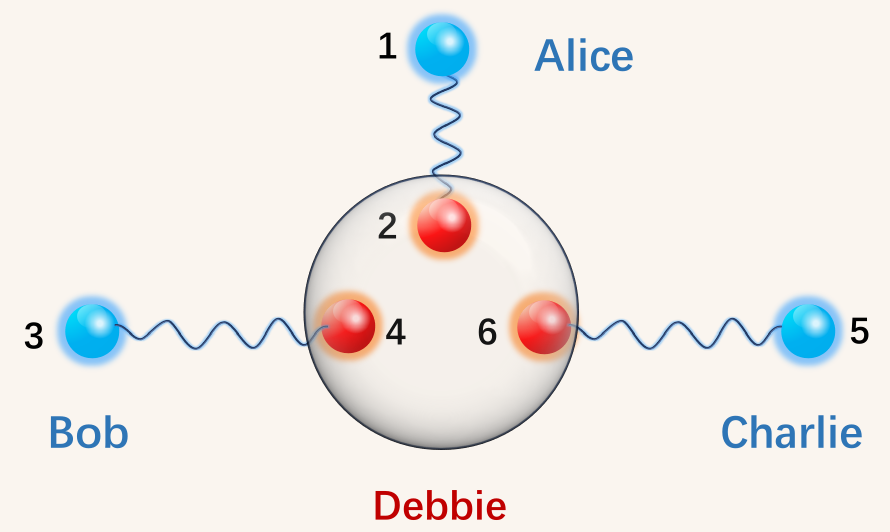} 
\caption{
Setup for 
demonstrating simultaneous refutation of local realism and noncontextuality.
Debbie starts with three singlet pairs and then sends
qubits of $1$, $3$, and $5$ to Alice, Bob, and Charlie, respectively. Each of
Alice, Bob, Charlie and Debbie is in a spacelike separated region from the
other three.
}%
\label{fig1}%
\end{figure}

Consider the arrangement shown in Fig.~\ref{fig1}. Initially, Debbie possesses three
pairs of antisymmetric singlet Bell states $|\Psi\rangle_{123456} = \left|
\psi^{-}\right\rangle _{12} \otimes\left| \psi^{-}\right\rangle _{34}%
\otimes\left| \psi^{-}\right\rangle _{56}$. Here the singlet state
\begin{equation}
\left\vert \psi^{-}\right\rangle _{ij}=\frac{1}{\sqrt{2}}(|01\rangle
_{ij}-|10\rangle_{ij}),
\end{equation}
where $|01\rangle_{ij}=|0\rangle_{i}\otimes|1\rangle_{j}$. Assume
Debbie keeps qubits $2$, $4$, and $6$ in her laboratory, and sends qubits
$1$, $3$, and $5$ to Alice, Bob, and Charlie, respectively. Each of these four
observers are in a spacelike separated region from the other three observers.
The qubits are then measured locally in their respective laboratories.
If Debbie performs the GHZ-state measurement, the arrangement considered here is
the GHZ-entanglement swapping as experimentally demonstrated in Ref.~\cite{lu}. \textcolor{black}{In our experimental setup, we assume that each pair of entangled sources is independent. This assumption is crucial, as it restricts our study to exploring quantum nonlocality within the specific context of network scenarios, where the independence of sources plays a key role in the manifestation of nonlocal correlations~\cite{vsupic2020quantum,cavalcanti2011quantum}.}

We first discuss the negation of local realism by our scheme. To make the following discussion clearly, we define $|\Psi\rangle=|\Psi\rangle_{123456}$,
$x_{i}=\sigma_{xi}$, $y_{i}=\sigma_{yi}$ ($i=1,2,...,6$) and $x_{2}y_{4}y_{6}=\sigma_{x2}\sigma_{y4}\sigma_{y6}$, etc., and apply $(\cdot)$ to separate local operators. Thus, one can easily check that the state $|\Psi\rangle$ satisfies the following eigenequations:

\begin{align}
&  \ \left.  x_{2}y_{4}y_{6}\cdot x_{1}\cdot y_{3}\cdot y_{5}\left\vert
\Psi\right\rangle =-\left\vert \Psi\right\rangle ,\right.  \label{e1}\\
&  \ \left.  y_{2}x_{4}y_{6}\cdot y_{1}\cdot x_{3}\cdot y_{5}\left\vert
\Psi\right\rangle =-\left\vert \Psi\right\rangle ,\right.  \label{e2}\\
&  \ \left.  y_{2}y_{4}x_{6}\cdot y_{1}\cdot y_{3}\cdot x_{5}\left\vert
\Psi\right\rangle =-\left\vert \Psi\right\rangle ,\right.  \label{e3}\\
&  \ \left.  x_{2}x_{4}x_{6}\cdot x_{1}\cdot x_{3}\cdot x_{5}\left\vert
\Psi\right\rangle =-\left\vert \Psi\right\rangle ,\right.  \label{e4}\\
&  \ \left.  x_{2}x_{4}x_{6}\cdot x_{2}y_{4}y_{6}\cdot y_{2}x_{4}y_{6}\cdot
y_{2}y_{4}x_{6}\left\vert \Psi\right\rangle =-\left\vert \Psi\right\rangle
.\right.  \label{e5}%
\end{align}
The last equation [Eq. (\ref{e5})] stems from the operator identity 
$x_{2}x_{4}x_{6}\cdot x_{2}y_{4}y_{6}\cdot y_{2}x_{4}y_{6}\cdot
y_{2}y_{4}x_{6}=-I_{246}$, where $I_{246}$ is the identity operator 
for three qubits $2$, $4$, and $6$.

Only local operators (Alice's, Bob's,
Charlie's, or Debbie's) are used in the equations (\ref{e1})-(\ref{e5}), implying that any three of Alice, Bob, Charlie and Debbie can confidently assign values to the local operators of the
remaining party by measuring their local observables without disturbing the
system of that party in any way. According to EPR's notion of local
realism~\cite{einstein1935can}, \textit{\textquotedblleft If, without in any
way disturbing a system, we can predict with certainty (i.e., with probability
equal to unity) the value of a physical quantity, then there exists an element
of physical reality corresponding to this physical quantity\textquotedblright%
}. As a result, according to those who believe in local realistic
theories, each local operator in Eqs.~(\ref{e1})-(\ref{e5}) separated by
$(\cdot)$ can be defined as EPR's local \textquotedblleft elements of
reality\textquotedblright. Moreover, all these elements of reality have
predetermined values, which are denoted by $v\left(  x_{i}\right)  $,
$v\left(  y_{i}\right)  $, $v\left(  x_{2}y_{4}y_{6}\right)  $, $v\left(
y_{2}x_{4}y_{6}\right)  $, $v\left(  y_{2}y_{4}x_{6}\right)  $, and $v\left(
x_{2}x_{4}x_{6}\right)  $ with $v=\pm1$. Therefore, Eqs.~(\ref{e1})-(\ref{e5})
can be rewritten in a local realistic theory as

\begin{align}
&  \ \left.  v(x_{2}y_{4}y_{6})v(x_{1})v(y_{3})v(x_{5})=-1,\right.
\label{e6}\\
&  \ \left.  v(y_{2}x_{4}y_{6})v(y_{1})v(x_{3})v(y_{5})=-1,\right. \\
&  \ \left.  v(y_{2}y_{4}x_{6})v(y_{1})v(y_{3})v(x_{5})=-1,\right. \\
&  \ \left.  v(x_{2}x_{4}x_{6})v(x_{1})v(x_{3})v(x_{5})=-1,\right. \\
&  \ \left.  v(x_{2}x_{4}x_{6})v(x_{2}y_{4}y_{6})v(y_{2}x_{4}y_{6}%
)v(y_{2}y_{4}x_{6})=-1.\right. \label{e10}%
\end{align}

However, it is simple to see that Eqs.~(\ref{e6})-(\ref{e10}) are mutually
incompatible. Because each $v$-value appears twice on the left-hand
sides of Eqs.~(\ref{e6})-(\ref{e10}), the product of all the left-hand sides
must equal $+1$, which contradicts the result $-1$ obtained by multiplying
all the right-hand sides. This contradiction demonstrates that any local realistic
theory based on EPR's notion of local realism is incapable of reproducing the results of
perfect correlations predicted by QM as in Eqs.~(\ref{e1})-(\ref{e5}). This brings the proof of all-versus-nothing nonlocality to a close. This type of logical impossibility proof is known as Bell's theorem without inequalities.
Our scheme's experimental issues (particularly the measurement strategy) will be discussed further below.

Bell's theorem without inequalities has been demonstrated for the
case of three~\cite{greenberger1989bell,greenberger1990bell,mermin1990pt,pan2000experimental} or
even two~\cite{cabello2001all,PhysRevLett.95.210401,chen2003all,yang2005all} spacelike separated observers. It is not surprising, then, that our argument using four spacelike
separated observers can demonstrate Bell's theorem without inequalities.
However, our argument differs from previous ones in that local realism and noncontextuality can be denied simultaneously by one and the same
experimental system. This type of argument sheds new light on the relationship between Bell's theorem and the KS theorem.

To deny noncontextuality in our scheme, we only need to focus on Debbie's
system of three qubits $2$, $4$, and $6$. Consider the ten operators $x_{2}$,
$y_{2}$, $x_{4}$, $y_{4}$, $x_{6}$, $y_{6}$, $x_{2}y_{4}y_{6}$, $y_{2}%
x_{4}y_{6}$, $y_{2}y_{4}x_{6}$, and $x_{2}x_{4}x_{6}$ of this system. Now we
have five operator identities~\cite{mermin1990simple} 

\begin{align}
&  \ \left.  x_{2}y_{4}y_{6}\cdot x_{2}\cdot y_{4}\cdot y_{6}=I_{246},\right.
\label{id1}\\
&  \ \left.  y_{2}x_{4}y_{6}\cdot y_{2}\cdot x_{4}\cdot y_{6}=I_{246},\right.
\label{id2}\\
&  \ \left.  y_{2}y_{4}x_{6}\cdot y_{2}\cdot y_{4}\cdot x_{6}=I_{246},\right.
\label{id3}\\
&  \ \left.  x_{2}x_{4}x_{6}\cdot x_{2}\cdot x_{4}\cdot x_{6}=I_{246},\right.
\label{id4}\\
&  \ \left.  x_{2}x_{4}x_{6}\cdot x_{2}y_{4}y_{6}\cdot y_{2}x_{4}y_{6}\cdot
y_{2}y_{4}x_{6}=-I_{246}.\right.  \label{id5}%
\end{align}

These identities are quantum mechanical predictions that are independent of the state of the system. Note that, in each of Eqs.~(\ref{id1})-(\ref{id5}), 
these operators separated by $(\cdot)$ are mutually commuting.
In an NCHV theory, these operators separated by $(\cdot)$ have predetermined
values, but here without appealing to EPR's local realism. Let us denote these
predetermined values by $v\left(  x_{i}\right)  $, $v\left(  y_{i}\right)  $,
$v\left(  x_{2}y_{4}y_{6}\right)  $, $v\left(
y_{2}x_{4}y_{6}\right)  $, $v\left(  y_{2}y_{4}x_{6}\right)  $, and $v\left(
x_{2}x_{4}x_{6}\right)  $
with $v=\pm1$. Therefore, Eqs.~(\ref{id1})-(\ref{id5}) can be rewritten in an NCHV theory as

\begin{align}
&  \ \left.  v(x_{2}y_{4}y_{6})v(x_{2})v(y_{4})v(y_{6})=1,\right.
\label{e11}\\
&  \ \left.  v(y_{2}x_{4}y_{6})v(y_{2})v(x_{4})v(y_{6})=1,\right.
\label{e12}\\
&  \ \left.  v(y_{2}y_{4}x_{6})v(y_{2})v(y_{4})v(x_{6})=1,\right.
\label{e13}\\
&  \ \left.  v(x_{2}x_{4}x_{6})v(x_{2})v(x_{4})v(x_{6})=1,\right.
\label{e14}\\
&  \ \left.  v(x_{2}x_{4}x_{6})v(x_{2}y_{4}y_{6})v(y_{2}x_{4}y_{6}
)v(y_{2}y_{4}x_{6})=-1.\right.  \label{e15}%
\end{align}

NCHV theories can explain the equations~(\ref{e11})-(\ref{e14}). According to NCHV theories, the predetermined values of these
operators can be measured without disturbance, regardless of
the context in which they are contained. In other words, no matter how an
operator $A$ is measured for a specific system, the result of measuring $A$
will always be $v\left(  A\right)  $. Therefore, $v(x_{2}y_{4}y_{6}%
)=v(x_{2})v(y_{4})v(x_{6})$ in Eq.~(\ref{e11}) because $x_{2}y_{4}y_{6}$ can
be measured by separately measuring $x_{2}$, $y_{4}$ and $y_{6}$ and
multiplying their results. Alternatively, because they are mutually commuting operators, $x_{2}y_{4}y_{6}$, $x_{2}$, $y_{4}$ and $y_{6}$ can be co-measured by
a single apparatus, yielding the product of the four operators to be $+1$. Likewise, Eqs.~(\ref{e12})-(\ref{e14}) are also
consequences of noncontextuality.  Finally, the left-hand side of 
Eq.~(\ref{e15}) is the product of mutually commuting 
operators $x_{2}y_{4}y_{6}$, $y_{2}x_{4}y_{6}$, $y_{2}y_{4}x_{6}$ and
$x_{2}x_{4}x_{6}$, which can be simultaneously measured by a single measurement apparatus, to be
considered below. Now, the contradiction between NCHV theories and QM can be seen by combining Eqs.~(\ref{e11})-(\ref{e15}). 

Because each $v$-value appears twice on the left-hand sides
of Eqs.~(\ref{e11})-(\ref{e15}) and can only be assigned the value $+1$ or
$-1$, the product of the left-hand sides must be $+1$, whereas the product of the right-hand sides must be $-1$. As a result, the
attempt to assign values to these ten operators, as shown in Eqs.~(\ref{e11})-(\ref{e15}), must fail. This
failure implies that any NCHV theory contradicts the quantum mechanical predictions in 
Eqs.~(\ref{id1})-(\ref{id5}). The contradiction occurs on a nonstatistical level in predictions for these operators. As a result, our scheme completes the demonstration of the all-versus-nothing 
contextuality without sepecifying any state of the system. Namely, our proof of contextuality is state-independent.

Note that Mermin~\cite{mermin1990simple}, generalizing a state-dependent argument due to Peres ~\cite{peres}, 
made use of the same set of operator identities in Eqs.~(\ref{id1})-(\ref{id5}) for his argument on the KS theorem. To validate his GHZ argument, the three qubits involved must be spacelike separated. In contrast, for both the KS and GHZ arguments, the three qubits are always kept 
on Debbie's side in our proof. Therefore, the current work represents a genuinely simultaneous all-versus-nothing refutation of local realism and noncontextuality by a single system.

In an ideal experimental situation, each of Eqs.~(\ref{e1})-(\ref{e5}) would be proven true by
experimental data when demonstrating the negation of local realism. This demonstrates that QM can withstand the experiment's scrutiny. Experiments would reject any local realistic theory that is
fundamentally contradictory to QM when combined with the preceding discussion.
Unfortunately, perfect correlations
and ideal measurement devices are difficult to prepare.
Therefore, to address these less-than-ideal experimental situations, one can introduce the operator
$\mathcal{O}=x_{2}y_{4}y_{6}\cdot x_{1}\cdot y_{3}\cdot y_{5}+y_{2}x_{4}%
y_{6}\cdot y_{1}\cdot x_{3}\cdot y_{5}+y_{2}y_{4}x_{6}\cdot y_{1}\cdot
y_{3}\cdot x_{5}+x_{2}x_{4}x_{6}\cdot x_{1}\cdot x_{3}\cdot x_{5}$, in which
$-x_{2}x_{4}x_{6}\cdot x_{2}y_{4}y_{6}\cdot y_{2}x_{4}y_{6}\cdot y_{2}%
y_{4}x_{6}$\ being an identity operator is not included. Combined with
Eqs.~(\ref{e1})-(\ref{e5}), one can easily check that $|\Psi\rangle$ is an
eigenstate of the operator $\mathcal{O}$, thus having
\begin{equation}
\mathcal{O}|\Psi\rangle=-4|\Psi\rangle.\label{quantum}%
\end{equation}
Because of the operator identity in Eq.~(\ref{id5}), one of the observed values 
of these four operators $x_2y_4y_6$, $y_2x_4y_6$, $y_2y_4x_6$ and $x_2x_4x_6$ is always 
different from the other three. As a result, any local realistic theory predicts the observed values of the
operator $\mathcal{O}$
\begin{equation}
\left\vert \langle\mathcal{O}\rangle_{LRT}\right\vert \leq2,\label{mermin}%
\end{equation}
which is a Bell-Mermin inequality~\cite{mermin1990extreme} for $|\Psi\rangle$. The inequality (\ref{mermin}) also
shows a contradiction with the prediction (\ref{quantum}) of QM, and can thus be
verified in a less-than-ideal experimental situation. Suppose that we have a noisy singlet 
state $F\left|\psi^{-}\right\rangle$$\left\langle\psi^{-}\right|+\frac{1-F}{4}I$, where $I$ is
the unit operator in the two-qubit state space. Then the 
violation of the inequality (\ref{mermin}) can be observed as long as the 
fidelity $F$ of the noisy singlet state is greater than $\sqrt[3]{1/2}\approx0.7937$. \textcolor{black}{While our experimental design is indeed based on logical contradictions to demonstrate the discrepancies between local realism and quantum mechanical predictions, it inherently does not negate the necessity for statistical analysis. In this context, the Bell-Mermin inequality links theory with the empirical evidence needed, highlighting the essential role of statistics in experimental physics to approximate theoretical ideals.}

Similarly, in an experimental demonstration of contextuality, 
one can define the operator
$\mathcal{O'}=x_{2}y_{4}y_{6}\cdot x_{2}\cdot y_{4}\cdot y_{6}+y_{2}x_{4}%
y_{6}\cdot y_{2}\cdot x_{4}\cdot y_{6}+y_{2}y_{4}x_{6}\cdot y_{2}\cdot
y_{4}\cdot x_{6}+x_{2}x_{4}x_{6}\cdot x_{2}\cdot x_{4}\cdot x_{6}$, for which
any NCHV theory must predict in a less-than-ideal experimental situation
\begin{equation}
\left\vert \langle\mathcal{O'}\rangle_{NCHV}\right\vert \leq2.\label{merminNCHV}%
\end{equation}
In contrast, the maximal value of quantum mechanical prediction 
of $\left\vert \langle\mathcal{O'}\rangle_{QM}\right\vert$ is 4. This difference 
allows a statistical test of NCHV theories in a less-than-ideal experimental situation. Suppose that we have a noisy singlet 
state $F\left|\psi^{-}\right\rangle$$\left\langle\psi^{-}\right|+\frac{1-F}{4}I$, where $I$ is
the unit operator in the two-qubit state space. Then the 
violation of the inequality (\ref{merminNCHV}) can be observed as long as the 
fidelity $F$ of the noisy singlet state is greater than $\sqrt[3]{1/2}\approx0.7937$.

We have discussed how our scheme demonstrates simultaneous refutation of local
realism and noncontextuality by one and the same experimental system in a
less-than-ideal experimental situation. Here we further emphasize that these
four operators $x_{2}y_{4}y_{6}$, $y_{2}x_{4}y_{6}$, $y_{2}y_{4}x_{6}$ and
$x_{2}x_{4}x_{6}$ are measured with one and the same apparatus, which is crucial for
demonstrating the negation of local realism~\cite{chen2003all}. In our
argument of the negation of local realism, each of the four operators
$x_{2}y_{4}y_{6}$, $y_{2}x_{4}y_{6}$, $y_{2}y_{4}x_{6}$ and $x_{2}x_{4}x_{6}$
appears twice in Eqs.~(\ref{e1})-(\ref{e5}). For example, the operator
$x_{2}y_{4}y_{6}$ appears in Eqs.~(\ref{e1}) and (\ref{e5}), whereas
Eqs.~(\ref{e1}) and (\ref{e5}) involve two different experimental contexts for measuring
$x_{2}y_{4}y_{6}$. LHV, with emphasis, does not assume noncontextuality. In other words, LHV only emphasizes that all operators have
predetermined values, but it does not rule out the possibility of measuring an operator causing a disturbance in the system, thereby modifying the values of
other operators. As a result, those who believe in local realistic theory can
argue that the contradiction in Eqs.~(\ref{e6})-(\ref{e10}) is caused by contextuality, rather than EPR's local elements of reality. To avoid this problem, one either adds the
additional assumption of noncontextuality, or measures the four operators $x_{2}%
y_{4}y_{6}$, $y_{2}x_{4}y_{6}$, $y_{2}y_{4}x_{6}$ and $x_{2}x_{4}x_{6}$ with
one and the same apparatus such that they are measured within a single context. Note that our argument on the negation of
noncontextuality does not require such apparatuses, since noncontextuality is
already assumed in this argument.

\begin{figure}[ptb]
\includegraphics[width=\columnwidth]{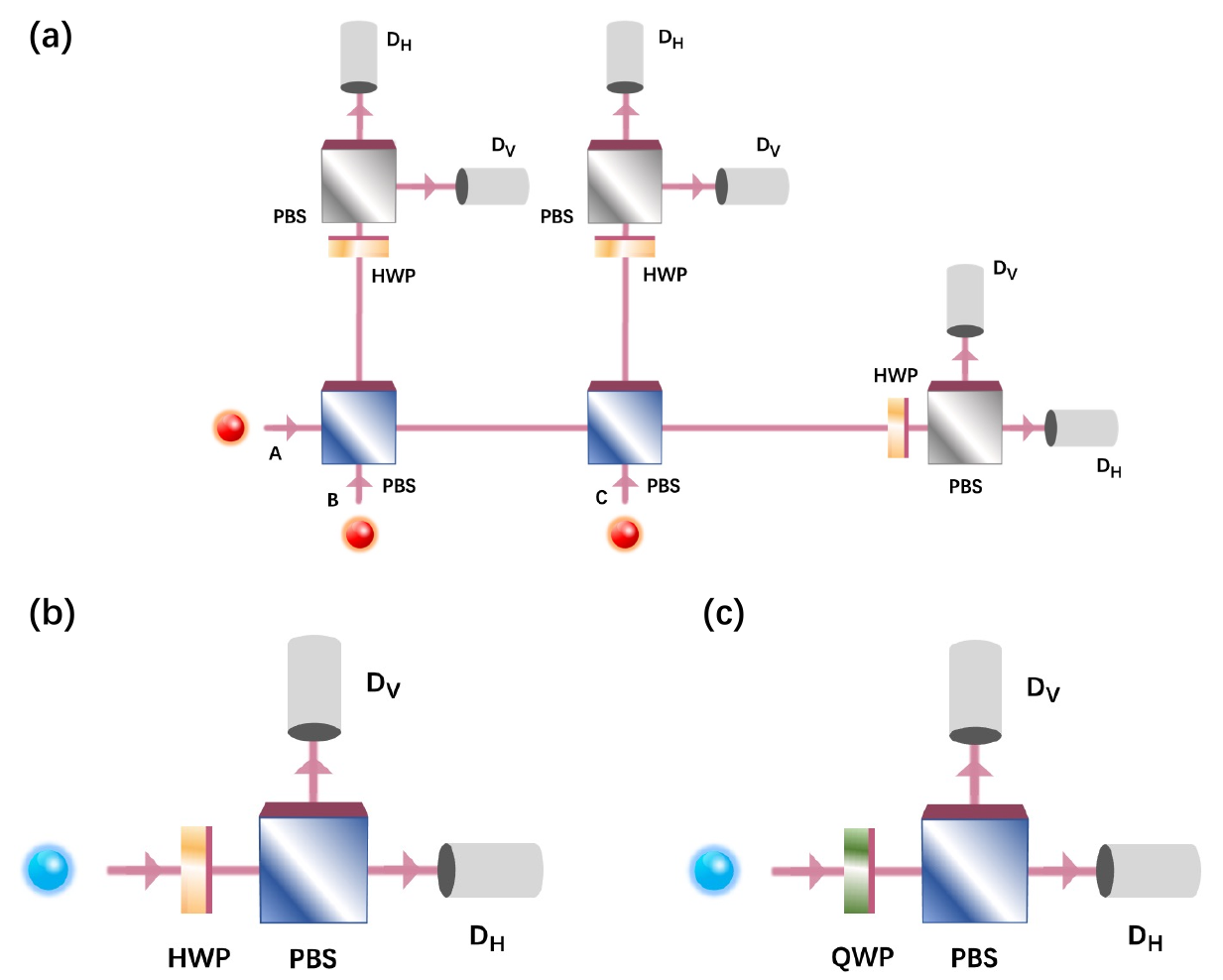} \caption{Devices for measuring
local operators in Eqs.~(\ref{e1})-(\ref{e5}). (a) A GHZ-state analyzer
realized with linear optics. Qubits $2$, $4$, and $6$ in Debbie are incident in
modes $A$, $B$, and $C$, respectively. The half-wave plate (HWP) axis is set at an angle of $22.5$ degrees relative to the horizontal direction, resulting in a $45$-degree rotation of the polarization. According to different threefold
coincidences recorded by detectors, one can judge whether the three photonic qubits
collapse to the state $\left| \Phi_{0}^{+}\right\rangle $ or the state $\left|
\Phi_{0}^{-}\right\rangle $, and thus simultaneously measure the values of the
four operators $x_{2}y_{4}y_{6}$, $y_{2}x_{4}y_{6}$, $y_{2}y_{4}x_{6}$ and
$x_{2}x_{4}x_{6}$; (b) Device for measuring $x_{1}$, $x_{3}$ and $x_{5}$. The angle between the HWP axis and the horizontal direction is set at $22.5$ degrees; (c)
Device for measuring $y_{1}$, $y_{3}$ and $y_{5}$. The angle between the quarter-wave plate (QWP) axis and the horizontal direction is set at $45$ degrees. PBS is polarizing beam
splitter, and $D_{H}$ and
$D_{V}$ are trigger detectors for horizontal ($H$) and vertical ($V$) polarizations, respectively. \textcolor{black}{The PBS is a critical component that separates the incoming light into two orthogonal polarization states, typically H and V.}}%
\label{fig2}%
\end{figure}

\section{CONCEPTUALIZATION OF AN EXPERIMENTAL SCHEME}

As a nonlocality argument, one of course has to avoid the introduction of 
additional assumption (i.e., noncontextuality). Then one needs to measure the four operators $x_{2}y_{4}y_{6}$, $y_{2}x_{4}y_{6}$,
$y_{2}y_{4}x_{6}$ and $x_{2}x_{4}x_{6}$ with the same apparatus. A possible
measurement strategy of our nonlocality argument is to perform the GHZ-state
measurement~\cite{pan1998greenberger} on photonic qubits $2$, $4$, and $6$ of Debbie
[Fig.~\ref{fig2}(a)]. With emphasis, photonic GHZ-state discrimination
realized with linear optics can only identify two out of the eight maximally
entangled GHZ states, namely, $\Phi_{0}^{\pm}=\frac{1}{\sqrt{2}}(|0\rangle
|0\rangle|0\rangle\pm|1\rangle|1\rangle|1\rangle)$, where $|0\rangle=|H\rangle$ 
for horizontal polarization and $|1\rangle=|V\rangle$ for vertical polarization. See Appendix A for a detailed discussion. Fortunately, this does not
affect our argument, because one only needs to postselect the corresponding
experimental data when $\Phi_{0}^{\pm}=\frac{1}{\sqrt{2}}(|0\rangle
|0\rangle|0\rangle\pm|1\rangle|1\rangle|1\rangle)$ is measured. To verify that
this post-selection is desirable and it does not affect our conclusions, one
can expand $\left\vert \Psi\right\rangle $ into the form of the GHZ-entanglement swapping~\cite{lu}

\begin{equation}
\begin{aligned} \left|\Psi\right\rangle &= \\ &\frac{1}{\sqrt{8}}\left(\Phi_{0(2,4,6)}^{-} \otimes \Phi_{0 (1,3,5)}^{+}-\Phi_{0(2,4,6)}^{+} \otimes \Phi_{0(1,3,5)}^{-}\right. \\ &+\Phi_{1(2,4,6)}^{+} \otimes \Phi_{1(1,3,5)}^{-}-\Phi_{1(2,4,6)}^{-} \otimes \Phi_{1(1,3,5)}^{+} \\ &+\Phi_{2(2,4,6)}^{+} \otimes \Phi_{2(1,3,5)}^{-}-\Phi_{2(2,4,6)}^{-} \otimes \Phi_{2(1,3,5)}^{+} \\ &\left.+\Phi_{3(2,4,6)}^{+} \otimes \Phi_{3(1,3,5)}^{-}-\Phi_{3(2,4,6)}^{-} \otimes \Phi_{3(1,3,5)}^{+}\right), \end{aligned}\label{ghz}%
\end{equation}
where $\Phi_{1}^{\pm}=\frac{1}{\sqrt{2}}(|1\rangle|0\rangle|0\rangle
\pm|0\rangle|1\rangle|1\rangle)$, $\Phi_{2}^{\pm}=\frac{1}{\sqrt{2}}%
(|0\rangle|1\rangle|0\rangle\pm|1\rangle|0\rangle|1\rangle)$, $\Phi_{3}^{\pm
}=\frac{1}{\sqrt{2}}(|0\rangle|0\rangle|1\rangle\pm|1\rangle|1\rangle
|0\rangle)$, and subscripts $(2,4,6)$ and $(1,3,5)$ represent qubits $2,4,6$
and qubits $1,3,5$, respectively. It can be checked that each component of the 
expansion in Eq.~(\ref{ghz}) satisfies Eqs.~(\ref{e1})-(\ref{e5}).
Thus, whenever a GHZ state, e.g., $\Phi_{0(2,4,6)}^{-}$, is successfully detected by Debbie, she simultaneously
gets the measurement values of the four operators, whose eigenstate is the detected GHZ state.
This property enables one to demonstrate our scheme using the GHZ post-selection
method and linear optics~\cite{fu,li,li2}, which is well within the reach of the current quantum
optical technology. The fact that the post-selection works here for our nonlocality argument
is not surprising as the successful GHZ detection by Debbie leaves qubits $1,3,5$ in an
``event-ready'' GHZ entanglement~\cite{lu}, which can of course be used in the GHZ argument.

\begin{figure}[ptb]
\includegraphics[width=\columnwidth]{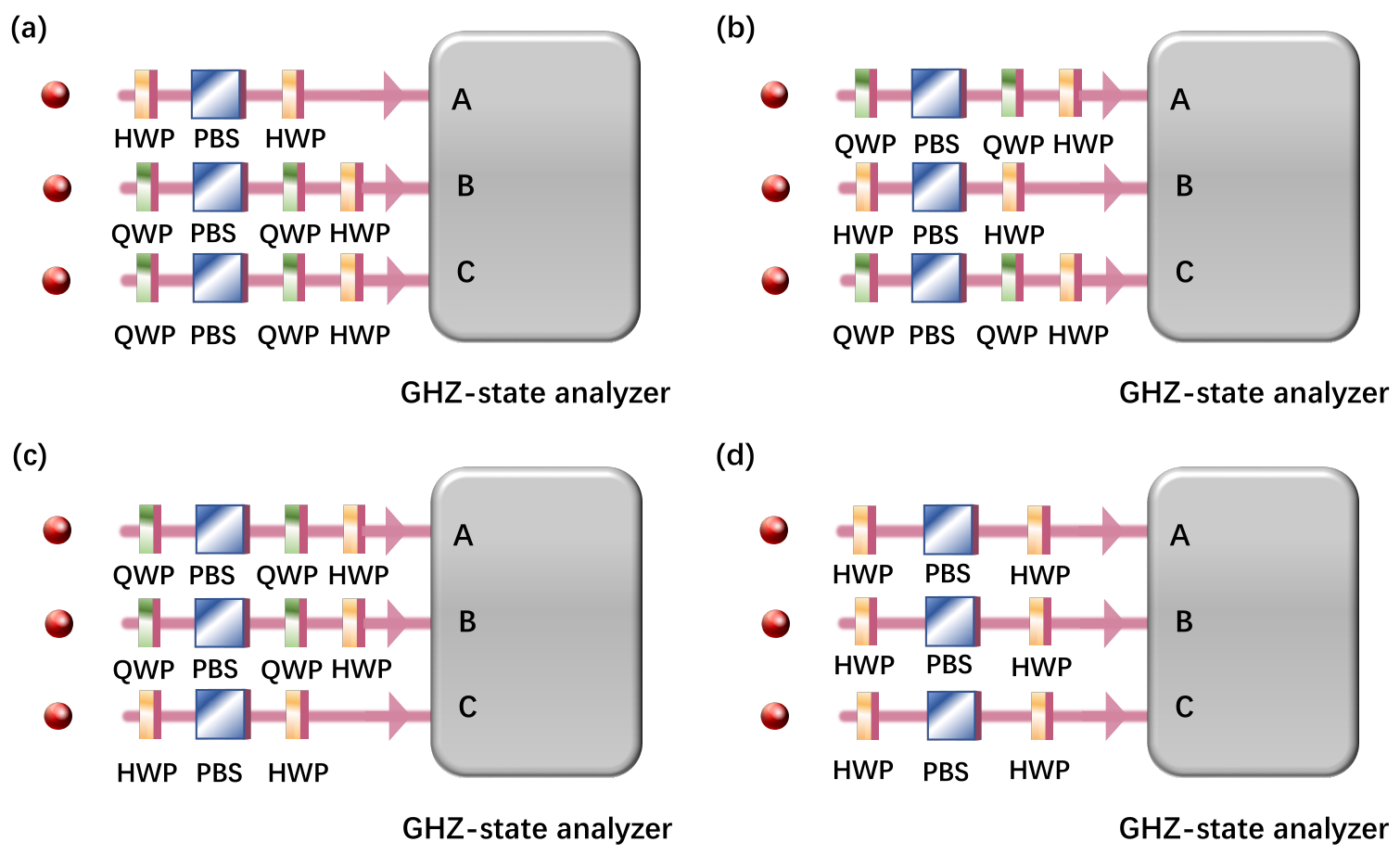} \caption{Devices for measuring
local operators in Eqs.~(\ref{id1})-(\ref{id4}). Photonic qubits $2$, $4$, and $6$ in Debbie are 
projected into definite eigenstates and then the incident in
modes $A$, $B$, and $C$, respectively. All HWPs are disposed with their axes at a $22.5$-degree angle relative to the horizontal direction, and all QWPs at a $45$-degree angle. Here the GHZ-state analyzer is the same as the experimental device shown in Fig.~\ref{fig2}(a).}
\label{fig3}%
\end{figure}

\begin{table}
\center
\caption{The results of jointly measuring $x_2x_4x_6$, $x_2y_4y_6$, $y_2x_4y_6$ and $y_2y_4x_6$ upon different GHZ states.}
\begin{tabular}
{c @{\hspace{0.4cm}} c @{\hspace{0.4cm}} c  @{\hspace{0.4cm}} c @{\hspace{0.4cm}} c }  \hline \hline
 & $x_2x_4x_6$ & $x_2y_4y_6$ & $y_2x_4y_6$ & $y_2y_4x_6$   \\ \hline
$|\Phi^+_0\rangle$ & +1 & -1 & -1 & -1  \\
$|\Phi^-_0\rangle$ & -1 & +1 & +1  & +1 \\
$|\Phi^+_1\rangle$ & +1 & -1 & +1 & +1 \\
$|\Phi^-_1\rangle$ & -1 & +1 & -1 & -1 \\
$|\Phi^+_2\rangle$ & +1 & +1 & -1 & +1 \\
$|\Phi^-_2\rangle$ & -1 & -1 & +1 & -1 \\
$|\Phi^+_3\rangle$ & +1 & +1 & +1 & -1 \\
$|\Phi^-_3\rangle$ & -1 & -1 & -1 & +1 \\
\hline \hline
\end{tabular}
\label{ghz-state}
\end{table}

Now let us consider the measurement strategy of our argument on the negation of noncontextuality. Note that there is no way to measure all $10$ operators in Eqs.~(\ref{id1})-(\ref{id5}) within a single context, as these operators are not all commutative. However, this does not pose a problem because noncontextuality is already assumed in the argument as stated above. To verify Eq.~(\ref{id5}), Debbie can measure therein the four operators $x_{2}y_{4}y_{6}$, $y_{2}x_{4}y_{6}$, $y_{2}y_{4}x_{6}$ and $x_{2}x_{4}x_{6}$ via the GHZ-state measurement as above. However, for
Eqs.~(\ref{id1})-(\ref{id4}) she has to adopt a different measurement strategy, namely, to test them in different measurement
contexts. One possible way to test Eq.~(\ref{id1}) is to first project $x_2$, $y_4$ and $y_6$ separately into their definite eigenstates, and then feed qubits $2$, $4$, and $6$ into the GHZ-state analyzer to measure $x_2y_4y_6$. Figure~\ref{fig3}(a) shows as an example the case where, before entering the GHZ-state analyzer, qubits $2$, $4$, and $6$ collapse to the state $|+RR\rangle$, with $|+\rangle$ ($|R\rangle$) being the eigenstate in the $X$ ($Y$) basis and subscripts omitted for simplicity. This means that the measurement results of $x_2$, $y_4$ and $y_6$ are all $+1$. Meanwhile, $|+RR\rangle$ can be expanded in terms of the GHZ states (see Appendix B)
\begin{equation}
|+RR\rangle = \frac{1}{2}(|\Phi^-_0\rangle+|\Phi^-_1\rangle+i|\Phi^+_2\rangle+i|\Phi^+_3\rangle).
\label{26}
\end{equation} 
When combined with Table~\ref{ghz-state} the joint measurement of $x_2y_4y_6$ for $|+RR\rangle$ must yield $+1$. As a result, as long as Debbie observes coincidence in the GHZ-state analyzer, Eq.~(\ref{id1}) is verified. With emphasis, this verification method is not $100\%$ successful because the probability of qubits $2$, $4$, and $6$ all entering the GHZ-state analyzer is only $1/8$, and the GHZ-state analyzer can only identify two of the eight maximally entangled GHZ states. Again, this does not affect our argument. Similarly, Fig.~\ref{fig3}(b-d) are used to verify Eqs.~(\ref{id2})-(\ref{id4}), respectively. Before entering the GHZ-state analyzer in Fig.~\ref{fig3}(b-d), qubits $2$, $4$, and $6$ collapse into $|R+R\rangle$, $|RR+\rangle$, and $|+++\rangle$, respectively. Here
\begin{equation}
\begin{array}{l}
|R+R\rangle = \frac{1}{2}(|\Phi^-_0\rangle+i|\Phi^+_1\rangle+|\Phi^-_2\rangle+i|\Phi^+_3\rangle) \\
|RR+\rangle = \frac{1}{2}(|\Phi^-_0\rangle+i|\Phi^+_1\rangle+i|\Phi^+_2\rangle+|\Phi^-_3\rangle) \\
|+++\rangle = \frac{1}{2}(|\Phi^+_0\rangle+|\Phi^+_1\rangle+|\Phi^+_2\rangle+|\Phi^+_3\rangle),
\end{array}
\label{27}
\end{equation}
which, together with Table~\ref{ghz-state}, confirm Eqs.~(\ref{id2})-(\ref{id4}). Thus, following the above
measurement strategy Eqs.~(\ref{id1})-(\ref{id5}) can be tested quantum mechanically, but are not consistent with NCHV.

\section{Conclusion}

Previous research has chiefly advanced our understanding of the intriguing connections between quantum contextuality and nonlocality, primarily focusing on the role of inequalities. Our study showcases the feasibility of demonstrating both quantum contextuality and nonlocality, devoid of inequalities, within a singular system. It is worth noting that Ref.~\cite{guo2018} illustrated the concurrent observation of quantum contextuality and nonlocality utilizing inequalities. Our approach, however, surpasses the strategy adopted in Ref.~\cite{guo2018} by eschewing the dependency on inequalities, echoing the superiority of the GHZ theorem over Bell’s theorem.

While Bell’s theorem utilizes statistical methods to challenge LHV, raising potential doubts regarding confidence levels, the GHZ theorem seeks to obliterate this statistical refutation, providing a conclusive rebuttal of LHV through singular, precise measurements under optimal conditions. Likewise, the methodology in Ref.~\cite{guo2018} observes a coincidence of classical and quantum theories at ``trivial points'' of perfect correlations, thereby instigating a hunt for discrepancies between QM and LHV (or NCHV) amidst a myriad of imperfectly correlated points, ultimately culminating in complex schemes. Moreover, this approach finds it challenging to pinpoint any contradictions between QM and LHV (or NCHV) for definite predictions.

In contrast, our technique significantly amplifies entanglement correlations by employing three pairs of antisymmetric singlet Bell states [Eqs.~(\ref{ghz})], thereby simplifying the framework to a great extent. The result is that we can demonstrate the contradiction between QM and LHV (or NCHV) even for certain predictions, namely, just a logical contradiction between a few equations is enough to reveal the contradiction between QM and LHV (or NCHV).

Furthermore, we have devised a complementary experimental scheme aligned with our theory, viable with current linear optics technology. Quantum nonlocality and contextuality are pivotal resources empowering various quantum information tasks to transcend classical boundaries. Nevertheless, leveraging them synergistically remains a complex issue. Our efforts offer new insights into this dilemma, with the hope of providing additional ideas for the development of novel quantum protocols. Future endeavours should pivot towards empirical validations to affirm the concurrent negation of local realism and noncontextuality within unified systems, along with an exploration into the fascinating interplay between these two quantum resources.

\textcolor{black}{In conclusion, we introduce an approach that demonstrates quantum contextuality and nonlocality without relying on inequalities, contrasting with traditional methods that use statistical analysis and inequalities. Our method employs three pairs of antisymmetric singlet Bell states and GHZ-state measurements, simplifying the experimental framework significantly. This approach allows for the clear demonstration of contradictions between quantum mechanics and local hidden variables, even with specific predictions. Our experimental scheme, compatible with current linear optics technology, offers new perspectives in quantum information science, highlighting the interplay between nonlocality and contextuality and their potential applications in quantum information tasks.}

\section*{CRediT authorship contribution statement}
\bf Min-Gang Zhou: \rm Writing–original draft, Visualization, Investigation.  \bf Hua-Lei Yin: \rm Writing–review editing, Validation, Supervision, Funding acquisition.   \bf Zeng-Bing Chen: \rm Writing–review editing, Validation, Supervision, Funding acquisition, Conceptualization.

\section*{Declaration of competing interest}
The authors declare that they have no known competing financial interests or personal relationships that could have appeared to influence the work reported in this paper.

\section*{Data availability}
No data was used for the research described in the article.

\section*{ACKNOWLEDGEMENTS}
We gratefully acknowledge the supports from the National Natural Science Foundation of China (No. 12274223), the Natural Science Foundation of Jiangsu Province (No. BK20211145), the Fundamental Research Funds for the Central Universities (No. 020414380182), the Key Research and Development Program of Nanjing Jiangbei New Area (No. ZDYD20210101),  the Program for Innovative Talents and Entrepreneurs in Jiangsu (No. JSSCRC2021484), and the Program of Song Shan Laboratory (Included in the management of Major  Science and Technology Program of Henan Province) (No. 221100210800-02).

\appendix
\section{IMPLEMENTATION AND ANALYSIS OF GHZ STATE}

Projecting photons into GHZ states is a necessary means for many quantum information tasks. However, due to the difficulty in implementing controlled-NOT (CNOT) gate required for GHZ state preparation through linear optics and single photons, current photonic GHZ-state analyzers probabilistically distinguish between two out of the $2^N$ maximally entangled GHZ states formed by $N$ particles.

In this work, photonic qubits $2$, $4$, and $6$ of Debbie are spectrally indistinguishable identical photons. Therefore, eight polarization GHZ states could be written as

\begin{equation}
\begin{array}{l}
\left|\Phi_0^{\pm}\right\rangle=\frac{1}{\sqrt{2}}(|0\rangle|0\rangle|0\rangle \pm|1\rangle|1\rangle|1\rangle), \\
\left|\Phi_1^{\pm}\right\rangle=\frac{1}{\sqrt{2}}(|1\rangle|0\rangle|0\rangle \pm|0\rangle|1\rangle|1\rangle), \\
\left|\Phi_2^{\pm}\right\rangle=\frac{1}{\sqrt{2}}(|0\rangle|1\rangle|0\rangle \pm|1\rangle|0\rangle|1\rangle), \\
\left|\Phi_3^{\pm}\right\rangle=\frac{1}{\sqrt{2}}(|0\rangle|0\rangle|1\rangle \pm|1\rangle|1\rangle|0\rangle) ,
\end{array}
\end{equation} where $|0\rangle=|H\rangle$ for horizontal polarization and $|1\rangle=|V\rangle$ for vertical polarization. The three photons are now fed into the GHZ analyzer by modes A, B, and C (Fig.~\ref{fig2}(a)). Since the PBSs transmit $H$ and reflect $V$ polarizations, only when all photons are transmitted or reflected can a coincidence detection at the three outputs be generated. The result is that two GHZ states, i.e. $\left|\Phi_0^{\pm}\right\rangle$, can be distinguished from these eight GHZ states. Further, an HWP can be placed after each PBS to distinguish $\left|\Phi_0^{\pm}\right\rangle$. In this case, the $\left|\Phi_0^{+}\right\rangle$ state will result in coincidences $+++$, $+--$, $-+-$, and $--+$, where $+/-$ is defined as $\pm 45^{\circ}$ linear polarizations. Correspondingly, the $\left|\Phi_0^{-}\right\rangle$ state will result in coincidences $---$, $-++$, $+-+$, and $++-$. In this way, photonic qubits $2$, $4$, and $6$ of Debbie are projected into one of the GHZ states and the corresponding physical values of $x_{2}y_{4}y_{6}$, $y_{2}x_{4}y_{6}$, $y_{2}y_{4}x_{6}$ and
$x_{2}x_{4}x_{6}$ are obtained.

\section{EXPANSION OF GHZ-STATES ON OTHER POLARIZATION BASIS}

The eight maximally entangled GHZ states formed by three photons can be expanded on other polarization basis. For example,

\normalsize{
\begin{equation}
\begin{aligned}
\left|\Phi_0^{+}\right\rangle & =\frac{1}{2}(|+++\rangle+|+--\rangle+|-+-\rangle+|--+\rangle) \\
& =\frac{1}{2}(|+R L\rangle+|+L R\rangle+|-R R\rangle+|-L L\rangle)\\
&=\frac{1}{2}(|R+L\rangle+|L+R\rangle+|R-R\rangle+|L-L\rangle) \\
& =\frac{1}{2}(|R L+\rangle+|L R+\rangle+|R R-\rangle+|L L-\rangle), \\
\end{aligned}
\end{equation}

\begin{equation}
\begin{aligned}
\left|\Phi_0^{-}\right\rangle & =\frac{1}{2}(|++-\rangle+|+-+\rangle+|-++\rangle+|---\rangle) \\
& =\frac{1}{2}(|+R R\rangle+|+L L\rangle+|-R L\rangle+|-L R\rangle)\\
&=\frac{1}{2}(|R+R\rangle+|L+L\rangle+|R-L\rangle+|L-R\rangle) \\
& =\frac{1}{2}(|R R+\rangle+|L L+\rangle+|R L-\rangle+|L R-\rangle),\\
\end{aligned}
\end{equation}

\begin{equation}
\begin{aligned}
\left|\Phi_1^{+}\right\rangle & =\frac{1}{2}(|+++\rangle+|+--\rangle-|-+-\rangle-|--+\rangle) \\
& =\frac{1}{2}(|+R L\rangle+|+L R\rangle-|-R R\rangle-|-L L\rangle)\\
&=\frac{1}{2 i}(|R+R\rangle-|L+L\rangle-|L-R\rangle+|R-L\rangle) \\
& =\frac{1}{2 i}(|R R+\rangle-|L L+\rangle-|L R-\rangle+|R L-\rangle), \\
\end{aligned}
\end{equation}

\begin{equation}
\begin{aligned}
\left|\Phi_1^{-}\right\rangle & =\frac{1}{2}(|++-\rangle+|+-+\rangle-|-++\rangle-|---\rangle) \\
& =\frac{1}{2}(|+R R\rangle+|+L L\rangle-|-R L\rangle-|-L R\rangle)\\
&=\frac{1}{2 i}(|R+L\rangle-|L+R\rangle+|R-R\rangle-|L-L\rangle) \\
& =\frac{1}{2 i}(|R L+\rangle-|L R+\rangle+|R R-\rangle-|L L-\rangle),\\
\end{aligned}
\end{equation}

\begin{equation}
\begin{aligned}
\left|\Phi_2^{+}\right\rangle & =\frac{1}{2}(|+++\rangle-|+--\rangle+|-+-\rangle-|--+\rangle) \\
& =\frac{1}{2 i}(|+R R\rangle-|+L L\rangle+|-R L\rangle-|-L R\rangle)\\
&=\frac{1}{2}(|R+L\rangle+|L+R\rangle-|R-R\rangle-|L-L\rangle) \\
& =\frac{1}{2 i}(|R R+\rangle-|L L+\rangle+|L R-\rangle-|R L-\rangle), \\
\end{aligned}
\end{equation}

\begin{equation}
\begin{aligned}
\left|\Phi_2^{-}\right\rangle & =\frac{1}{2}(|++-\rangle-|+-+\rangle+|-++\rangle-|---\rangle) \\
& =\frac{1}{2 i}(|+R L\rangle-|+L R\rangle+|-R R\rangle-|-L L\rangle)\\
&=\frac{1}{2}(|R+R\rangle+|L+L\rangle-|R-L\rangle-|L-R\rangle) \\
& =\frac{1}{2 i}(|L R+\rangle-|R L+\rangle+|R R-\rangle-|L L-\rangle),\\
\end{aligned}
\end{equation}

\begin{equation}
\begin{aligned}
\left|\Phi_3^{+}\right\rangle & =\frac{1}{2}(|+++\rangle-|+--\rangle-|-+-\rangle+|--+\rangle) \\
& =\frac{1}{2 i}(|+R R\rangle-|+L L\rangle+|-L R\rangle-|-R L\rangle)\\
&=\frac{1}{2 i}(|R+R\rangle-|L+L\rangle+|L-R\rangle-|R-L\rangle) \\
& =\frac{1}{2}(|R L+\rangle+|L R+\rangle-|R R-\rangle-|L L-\rangle), \\
\end{aligned}
\end{equation}

\begin{equation}
\begin{aligned}
\left|\Phi_3^{-}\right\rangle & =\frac{1}{2}(|+-+\rangle-|++-\rangle+|-++\rangle-|---\rangle) \\
& =\frac{1}{2 i}(|+L R\rangle-|+R L\rangle+|-R R\rangle-|-L L\rangle)\\
&=\frac{1}{2 i}(|L+R\rangle-|R+L\rangle+|R-R\rangle-|L-L\rangle) \\
& =\frac{1}{2}(|R R+\rangle+|L L+\rangle-|R L-\rangle-|L R-\rangle),
\end{aligned}
\end{equation}}where $|+\rangle$ and $|-\rangle$ ($|R\rangle$ and $|L\rangle$) are the eigenstates in the $X$ ($Y$) basis. The Eqs.~\ref{26}, Eqs.~\ref{27} and Table~\ref{ghz-state} can be calculated according to the above formulas, thus the feasibility of the scheme in Fig.~\ref{fig3} can be verified.

\bigskip



\end{document}